\documentclass[twocolumn,showpacs,preprintnumbers,amsmath,amssymb]{revtex4}

\usepackage{epsfig,psfrag}
\usepackage{dcolumn}
\usepackage{bm}
\usepackage{graphicx}
\usepackage{color}

\begin{document}
\title{Four-body problem and BEC-BCS crossover 
in a quasi-one-dimensional cold fermion gas}
\author{C.~Mora,$^1$ A.~Komnik,$^2$ R.~Egger,$^1$ and A.O.~Gogolin$^3$ }
\affiliation{${}^1$~Institut f\"ur Theoretische Physik, 
Heinrich-Heine-Universit\"at,
 D-40225 D\"usseldorf, Germany\\
${}^2$  Physikalisches Institut, Albert-Ludwigs-Universit\"at, D-79104 
Freiburg, Germany\\
${}^3$ Department of Mathematics, Imperial College London, 180 Queen's Gate,
London SW7 2BZ, United Kingdom 
}
\date{\today}
\begin{abstract}
The four-body problem for an interacting two-species Fermi gas 
is solved analytically in a 
confined quasi-one-dimensional geometry, where the two-body atom-atom
scattering length $a_{aa}$ displays a confinement-induced resonance.
We compute the dimer-dimer scattering length $a_{dd}$, and show that this
quantity completely determines the many-body solution of the associated
BEC-BCS crossover phenomenon in terms of bosonic dimers.  
\end{abstract}

\maketitle
Cold atomic quantum gases continue to attract a lot of attention due to
their high degree of control, tunability, and versatility.
A main topic of interest has been the exploration of the 
BEC-BCS crossover in fermionic systems 
\cite{molecule1,regal04,grimm04,ketterle04,bourdel,chin}. 
In two or three dimensions, this is still a controversial
and not completely settled
issue on the theory side \cite{randeria,timmermanns,stoof,strinati},  
despite the qualitative agreement between mean-field theories and
 experimental data.  Notably,  a similar (but different) crossover phenomenon
has been predicted to occur in quasi-one-dimensional (1D) systems 
\cite{tokatly,fuchs}, where a cylindrical trap leads to a confinement-induced
resonance (CIR) \cite{olshanii98,bergeman03}
in the atom-atom interaction strength, analogous
to the magnetically tuned Feshbach resonance \cite{timmermanns}.
In contrast to what happens in 3D,  one {\sl always} has a 
two-body bound state (`dimer') in 1D, regardless
of the sign of the 3D atom-atom scattering length $a$. 
We solve the fermionic four-body problem in the confined
geometry, and compute the dimer-dimer scattering length
$a_{dd}$ throughout the full BCS-BEC crossover, on each side
of the CIR. 
On the `BEC' side, we establish 
contact to results for the unconfined case \cite{petrov04},
while on the `BCS' side,  a simple Bethe Ansatz calculation
 provides exact results.  The three-body problem has no trimer solution
\cite{mora}, and thus the {\sl full many-body crossover solution
can be expressed in terms of} $a_{dd}$ alone and is thereby
solved completely in this Letter. 
Since 1D atomic gases can be prepared and probed thanks to recent advances
\cite{gorlitz,esslinger,paredes},
our predictions could be observed in state-of-the-art experiments. 

We assume two fermion hyperfine components (denoted by $\uparrow,\downarrow$) 
with identical particle numbers
$N_\uparrow=N_\downarrow=N/2$, interacting only via $s$-wave
interactions.  
At low energies, the pseudopotential approximation \cite{huang} for
the 3D interaction among unlike fermions applies, 
$V({\bf r})=(4\pi\hbar^2 a/m_0) \delta({\bf r})
\partial_r (r\cdot)$ ($m_0$ is the mass).
We consider the  transverse confinement potential $U_c({\bf r})=
m_0 \omega_\perp^2 (x^2+y^2)/2$, with lengthscale $a_\perp=(2
\hbar/m_0\omega_\perp)^{1/2}$.  The solution of the two-body problem
\cite{olshanii98,bergeman03} reveals that a single
dimer (composite boson) state exists for arbitrary $a$, where
the dimensionless binding energy $\Omega_B$ and
 (longitudinal) size $a_B$,
\begin{equation}\label{omegab}
 \Omega_B=- \frac{E_B}{2 \hbar \omega_\perp} = 
 (a_\perp/2a_B)^2 > 0,
\end{equation}
follow from $\zeta(1/2,\Omega_B)=- a_\perp/a$ with 
the Hurvitz zeta function. For an experimental verification, see
Ref.~\cite{esslinger}.
For $a_\perp/a\to -\infty$, the BCS limit with $\Omega_B\simeq (a/a_\perp)^2
\ll 1$ and  $a_B\simeq a_\perp^2/2 |a|$
 is reached, while for $a_\perp/a\to +\infty$, the dimer (or BEC) limit emerges,
with $\Omega_B\simeq (a_\perp/2a)^2\gg 1$ and
 $a_B\simeq a$.  
The atom-atom scattering length is
\begin{equation}\label{aaa}
a_{aa}=a_\perp({\cal C}-a_\perp/a)/2, \quad
{\cal C}=-\zeta(1/2)\simeq 1.4603.
\end{equation}
For low energies, this result is reproduced by the 1D 
atom-atom interaction $V_{aa}(z,z')=g_{aa}\delta(z-z')$ with 
$g_{aa}=-2\hbar^2/m_0 a_{aa}$ \cite{olshanii98}.
The CIR (where $g_{aa}\to \pm \infty$)
takes place for $a_\perp/a={\cal C}$, which is equivalent to $\Omega_B=1$.  
In this paper, we solve the 1D fermionic four-body ($\uparrow
\uparrow\downarrow\downarrow$) problem and show that this also solves
the $N$-body problem for arbitrary $\Omega_B$ in the low-energy regime.

Let us first discuss general symmetries of the four-body problem.
We denote the positions of the $\uparrow$ ($\downarrow$) fermions
by ${\bf x}_{1,4}$ (${\bf x}_{2,3}$), respectively,
and then form distance vectors between unlike fermions,
${\bf r}_1={\bf x}_1-{\bf x}_2, {\bf r}_2={\bf x}_4-{\bf x}_3,$ and 
${\bf r}_+={\bf x}_1 -{\bf x}_3, {\bf r}_-={\bf x}_4 -{\bf x}_2.$
The distance vector between dimers $\{ 12 \}$ and $\{ 34 \}$ is 
${\bf R}/\sqrt{2}= ({\bf x}_1+{\bf x}_2-{\bf x}_3-{\bf x}_4)/2$.
After an orthogonal transformation, the center-of-mass coordinate
decouples and the four-body wavefunction $\Psi$ depends 
only on ${\bf r}_{1,2}$
and ${\bf R}$. With respect to dimer interchange,  $\Psi$ is symmetric,
\begin{equation}  \label{symmetryrelation}
 \Psi({\bf r}_1, {\bf r}_2, {\bf R}) = \Psi({\bf r}_2,{\bf r}_1,-{\bf R}) ,
\end{equation}
while under the exchange of identical fermions,
\begin{equation} \label{asymmetryrelation}
 \Psi({\bf r}_1,{\bf r}_2,{\bf R}) = -
\Psi({\bf r}_\pm, {\bf r}_\mp, \pm({\bf r}_1-{\bf r}_2)/\sqrt{2})  .
\end{equation}
The four-body Schr\"odinger equation then reads 
\begin{equation}     \label{4bodySE}
\begin{split}
 &  \Bigl [ - \frac{\hbar^2}{m_0} (\Delta_{{\bf r}_1} 
+ \Delta_{{\bf r}_2} + \Delta_{\bf R})
 + U_c({\bf r}_1) + U_c({\bf r}_2) \\ 
 & +  U_c({\bf R}) + V({\bf r}_2) -  E\Bigr] \Psi
  = - \sum_{i=1,\pm} V({\bf r}_i) \Psi.
\end{split}
\end{equation}
The pseudopotentials on the r.h.s. 
are incorporated  via Bethe-Peierls boundary conditions imposed when
a dimer is contracted, e.g.,
\begin{equation}   \label{BC1}
 \Psi({\bf r}_1,{\bf r}_2,{\bf R})|_{{\bf r}_1 \to 0 }
\simeq \frac{f({\bf r}_2,{\bf R})}{4 \pi r_1} ( 1 - r_1/a ) .
\end{equation}
All other boundary conditions can also be expressed in terms of
$f({\bf r},{\bf R})$ using Eqs.~(\ref{symmetryrelation}) 
and (\ref{asymmetryrelation}), where 
\begin{equation}\label{fsym}
f({\bf r},{\bf R})=f(-{\bf r},-{\bf R}),
\end{equation}
expresses (parity) invariance of Eq.~(\ref{4bodySE}) 
under ${\bf r}_{1,2}\to -{\bf r}_{1,2}$ and  ${\bf R}\to -{\bf R}$
in combination with Eq.~(\ref{symmetryrelation}). 
In order to appreciate the importance of Eq.~(\ref{fsym}),
it is instructive to expand $f({\bf r},{\bf R})$ in terms of
the single-particle eigenfunctions $\psi_\lambda ({\bf r})$ and
the two-body scattering states $\Phi_\lambda ({\bf r})$ in the 
presence of the confinement,
\begin{equation}\label{expansion}
f( {\bf r},{\bf R}) = \sum_{\mu \nu} f_{\mu \nu}
\Phi_\mu({\bf r}) \psi_\nu({\bf R}).
\end{equation}
The quantum numbers $\lambda$ include the 1D momentum $k$ \cite{footnote}, the
(integer) angular momentum $m$, and the radial quantum number $n=0,1,2,\ldots$.
Explicit expressions for $\psi_\lambda$ and $\Phi_\lambda$ can be found in 
Refs.~\cite{olshanii98,mora}. While both have
the same energy $E_{\lambda}$, the $\Phi_\lambda$ now include
the dimer bound state (denoted by $\lambda=0$) $\Phi_0({\bf r})$.
For relative longitudinal momentum $k$ of the two dimers,
the total energy is (excluding zero-point and center-of-mass motion) 
\begin{equation}\label{energy1}
E = - 2 \hbar \omega_\perp \Omega_B +\frac{ \hbar^2 k^2}{2 m_0}.
\end{equation}
We consider the low-energy regime $k a_\perp<1$,
where the relative dimer motion is in the lowest transverse state ($n=m=0$)
when dimers are far apart.  We then have to deal with a 1D 
dimer-dimer scattering problem in this `open' channel, where
the asymptotic {\sl 1D scattering state} $f_0(Z)$ 
for $|Z| \gg {\rm max} (a_\perp,|a_{aa}|)$
follows from Eq.~(\ref{expansion}) as
\begin{equation}\label{sccc}
f({\bf r},{\bf R})=\Phi_0({\bf r}) \psi_{\perp,00}\left
(\sqrt{X^2+Y^2}\right) f_0(Z),
\end{equation}
where $\psi_{\perp,00}$ is the transverse part of $\psi_{n=0,m=0}$.
The symmetry relation (\ref{fsym}) now enforces $f_0(Z)=f_0(-Z)$, 
reflecting the fact that two (composite) {\sl bosons} collide, i.e.,
\begin{equation}\label{asymptof}
f_0(Z) =  e^{-i k |Z|} + ( 1+ 2 \tilde{f}(k) )
e^{ik |Z|}.
\end{equation}
As long as only $s$-wave scattering is important, 
symmetry considerations thus rule out odd-parity solutions 
normally present in 1D scattering problems \cite{olshanii98,mora}.  
This crucial observation implies that, 
assuming analyticity, the 1D scattering amplitude can be expanded 
in terms of a {\sl 1D dimer-dimer scattering length}\ $a_{dd}$
\cite{landau},
\begin{equation}\label{expa}
\tilde{f}(k) = -1 + ik a_{dd} + {\cal O}(k^2).
\end{equation}
For $|k a_{dd}| \ll 1$, this also
follows from the zero-range 1D dimer-dimer potential
\begin{equation}\label{1dd}
V_{dd}(Z,Z')= g_{dd} \delta(Z-Z'), \quad g_{dd}=
-\frac{2 \hbar^2}{(2 m_0) a_{dd}}.
\end{equation}
We stress that Eq.~(\ref{1dd}) holds for arbitrary $a_\perp/a$, and 
therefore 1D dimer-dimer scattering at low energies is 
always characterized by a simple $\delta$-interaction.

Let us then analyze the {\sl BCS limit}, $\Omega_B\ll 1$,
where the scattering problem is kinematically 1D on lengthscales exceeding
$a_\perp$.  Projecting Eq.~(\ref{4bodySE}) onto the 
transverse ground state, the 1D Schr\"odinger equation for four 
attractively interacting fermions reads
with $a_{aa}=a_\perp^2/2 |a|\gg a_\perp$, see Eq.~(\ref{aaa}),
\begin{equation}\label{s2}
 \left( \frac{2 m_0 E}{\hbar^2}
 + \sum_{i=1}^4 \partial_{z_i}^2 + \frac{4}{a_{aa}} \sum_{i<j}
\delta(z_i-z_j) \right) \Psi=0,
\end{equation}
where the second sum excludes identical fermion pairs, 
$(i,j)$ corresponding to $\{14\}$ and $\{23\}$.
The Bethe Ansatz expresses the wavefunction as a sum of products of plane
waves \cite{models}. Let us choose the momenta
$a_{aa} k_{1,4} = \mp i - u /2$ and
$a_{aa} k_{3,2} = \mp i +u/2$
to describe dimer-dimer scattering,
and measure lengths in units of $a_{aa}$. 
The energy of this state is
$E =\hbar^2(-2+u^2/2)/(m_0  a_{aa}^2)$ and $u$ the relative
momentum of the two dimers.
Up to an overall normalization constant,
the wavefunction in the domain
${\cal D}_1 = \{ (z_1,z_4) < (z_3,z_2) \}$ 
must then be given by
\begin{eqnarray*}
\Psi_1  & = & e^{-(z_2+z_3-z_4-z_1)} \Bigl( 
e^{iu(z_2+z_4-z_3-z_1)/2} \\
&-& e^{iu(z_2+z_1-z_3-z_4)/2} + 
e^{iu (z_3+z_1-z_2-z_4)/2} \\
&-& e^{iu(z_3+z_4-z_2-z_1)/2} \Bigr)
\end{eqnarray*}
to ensure a normalizable and  antisymmetric solution
under exchange of identical fermions.
Consider next a second domain, ${\cal D}_2 = \{ z_1 < z_3 < z_4 < z_2\}$,
 where $z_3$ and $z_4$ are exchanged compared to ${\cal D}_1$. 
At the boundary between ${\cal D}_1$ and ${\cal D}_2$, $z_3=z_4$,
Eq.~(\ref{s2}) implies $\Psi_1=\Psi_2$ and
$(\partial_{z_3}-\partial_{z_4})(\Psi_1-\Psi_2)=-4\Psi_1$ \cite{lieb2}, 
leading to
\begin{eqnarray*}
\Psi_2 & =& 2 {\rm Re} \Big [
e^{-(z_2+z_3-z_4-z_1)} 
\frac{iu}{2+iu} e^{iu(z_2+z_4-z_3-z_1)/2} \\ 
&+& e^{-(z_2+z_4-z_3-z_1)} \Bigl(
 \frac{2}{2+iu} e^{iu (z_2+z_3-z_4-z_1)/2} \\ &-&
2 e^{i u(z_2+z_1-z_4-z_3)/2)} \Bigr)\Big ] .
\end{eqnarray*}
The wavefunction in other domains can be found in a similar manner.
As a result, for a large dimer-dimer distance $Z$,
 $\Psi \propto e^{-|z_+|} e^{-|z_-|} f_0(Z)$, where
$e^{-|z_{\pm}|}$ is the 1D wavefunction of the dimer $\{13\}$ and
$\{24\}$, respectively. 
This result shows explicitly that even in the BCS limit, the {\sl 
two dimers are
not broken} in the collision even for large $k$. 
There is no coupling to additional fermionic states, and
the {\sl composite} nature of the dimer is not apparent in $\Psi$.
The 1D scattering state $f_0(Z)$,
see Eq.~\eqref{asymptof}, has the {\sl exact}\ scattering amplitude
\begin{equation}\label{tfres}
\tilde{f} (k) = - \frac{1}{1+i k a_{dd}},\quad
a_{dd}=\frac{a_{aa}}{2}=\frac{ a_\perp^2}{4|a|},
\end{equation}
which reproduces the full scattering amplitude 
derived from Eq.~(\ref{1dd}) and not just the first order as in
Eq.~(\ref{expa}).
The bound state at imaginary $k$ predicted by Eq.~(\ref{tfres})
is however unphysical, since the corresponding Bethe Ansatz solutions are
then not normalizable. It would correspond to a
non-existent bound four-fermion (tetramer) state, and hence  Eq.~(\ref{tfres}) 
is restricted to the real axis.

Let us now turn to the many-body problem, starting with the BCS limit.
Since dimers are not broken in the collision, the ground state can be described
in terms of $N/2$ bosons (`bosonization') with
the interaction (\ref{1dd}) and
$a_{dd} = a_{aa}/2$.  The attractively interacting Bose gas is 
stabilized by the real-$k$ restriction, implying
the omission of many-body bosonic bound states.  Bosonization is possible for
$\rho \, a_\perp<1$, since typical momenta are $k \approx \rho$
for total 1D fermionic density $\rho$.  
This reasoning immediately leads to the famous
Lieb-Liniger (LL) equations \cite{lieb}, 
\begin{subequations}\label{lieb}
\begin{align}
\frac{E_0}{N} & =  -\hbar\omega_\perp \Omega_B
 + \frac{1}{\rho} \int_{-K_0}^{K_0} d k 
\frac{\hbar^2 k^2}{4 m_0} f(k), \\
2 \pi f(k) & =  1 - \frac{4}{a_{dd}} \int_{-K_0}^{K_0} d p \, 
\frac{f(p)}{4/a_{dd}^2 + (p-k)^2},
\end{align}
\end{subequations}
where $E_0$ is the ground state energy and $K_0$ is fixed by 
$\rho/2 = \int_{-K_0}^{K_0} d k f(k)$.
Notably, since $a_{dd} = a_{aa}/2$, the LL
equations coincide with Yang-Gaudin
equations for $N$ attractively interacting $1$D fermions, 
thereby explaining a deep connection noticed previously 
\cite{tokatly,fuchs,gaudin}. 
Moving towards the dimer limit, Eq.~(\ref{1dd}) still applies, but now only
for sufficiently small $k$ such that Eq.~(\ref{expa}) holds,
and  $a_{dd} \ne a_{aa}/2$.
For $a_{dd}\alt a_\perp$, one leaves the BCS regime and enters
the `crossover regime', while (once $a_{dd}<0$)
the dimer regime is realized for $|a_{dd}|\agt a_\perp$.  
Within the crossover regime, $|a_{dd}|\alt a_\perp$,
we have hard-core bosons that can effectively be fermionized 
\cite{tokatly,fuchs}, again implying typical momenta 
$k \approx \rho$. For  $\rho \, a_\perp <1$,
the condition $|k a_{dd}| \ll 1$ imposed by Eq.~(\ref{expa}) is
therefore safely fulfilled throughout the crossover regime.  
Finally, in the {\sl dimer limit}, $a<a_\perp$, 
fermions form very tightly bound dimers. The confinement can then
not influence the four-body collision, which is therefore
described by a 3D zero-range interaction with 
$a_{dd}^{3D}\approx 0.6 a$ \cite{petrov04}. 
However, for dimer-dimer distance larger than $a_\perp$,
 dimers eventually must occupy the 
transverse ground state, see Eq.~(\ref{sccc}).  In effect,
for $\rho \, a_\perp<1$, we recover a 1D (bosonic)  two-body problem, where
Eq.~(\ref{aaa}) gives the answer (exact for $\Omega_B\gg 1$), 
\begin{equation}\label{dimerlimit}
a_{dd} = - \frac{ a_{\rm red,\perp}^2}{2 (0.6 a) } , \quad
a_{\rm red,\perp}=(\hbar^2/m_0 \omega_\perp)^{1/2},
\end{equation}
where $a_{\rm red,\perp}$ is the transverse 
oscillator length for dimers.
To summarize this discussion, we have shown that (a)
as long as the single condition $k a_\perp<1$  holds,
dimer-dimer scattering is described by Eq.~(\ref{1dd})
for {\sl arbitrary} $a_\perp/a$, and (b)
knowledge of $a_{dd}$ and hence the solution
of the 1D four-body problem is sufficient to completely solve 
the 1D BEC-BCS many-body problem for dilute systems, $\rho \, a_\perp<1$,
in terms of the LL equations (\ref{lieb}).

\begin{figure}[t!]
\scalebox{0.3}{
\includegraphics{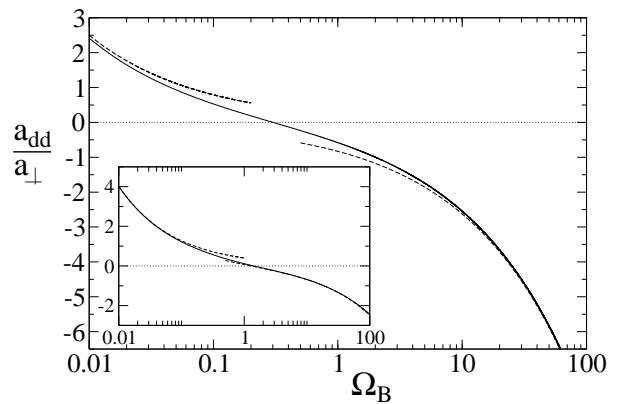} }
\caption{\label{f1}
Scattering length $a_{dd}$ as a function of $\Omega_B$.  
Dashed curves give exact limiting results,  
the solid curve interpolates by adding these.
\emph{Inset:}  Same but neglecting all closed-channel excitations. 
Here the solid curve gives the exact result.
}
\end{figure}

Next we discuss the {\sl 1D four-body problem}.
Enforcing the boundary condition (\ref{BC1}) or the other equivalent
ones, Eq.~(\ref{4bodySE}) leads to an integral equation
for  $f({\bf r},{\bf R})$ \cite{petrov04,mora}.
Using Eq.~(\ref{expansion}), some algebra \cite{mora3} yields 
\begin{eqnarray}\label{basiceq}
&& \left [\zeta\left(1/2,
\frac{E_\mu+E_\nu-E}{2\hbar\omega_\perp}
\right)-\zeta(1/2,\Omega_B)\right] f_{\mu\nu} \\ \nonumber
&& =\frac{4\pi \hbar^2 a_\perp}{\sqrt{2} m_0} \sum_{\mu^\prime \nu^\prime}
{\cal G}^{\mu^\prime \nu^\prime}_{\mu\nu} f_{\mu^\prime \nu^\prime} \, , \\
\nonumber && {\cal G}^{\mu^\prime \nu^\prime}_{\mu\nu} = 
\sum_\pm \int \, d{\bf r} \, d{\bf R}\,
 G_{E-E_\mu-E_\nu}( ({\bf r}\pm \sqrt{2} {\bf R})/2,0)\\ \nonumber
&&  \Phi^*_{\mu}\left(\frac{{\bf r} \mp \sqrt{2} {\bf
 R}}{2}\right)\psi^*_{\nu}\left(\mp\frac{{\bf r}}{\sqrt{2}}\right)
  \Phi_{\mu^\prime}({\bf r}) \psi_{\nu^\prime}({\bf R}).
\end{eqnarray}
The two-body Green's function $G_E({\bf r},0)$ can be found in 
Ref.~\cite{mora}.
The two degrees of freedom in $f({\bf r}, {\bf R})$ imply
two different types of `closed' channels 
that may be excited in a dimer-dimer collision:
(i) {\sl scattering states} above the bound state for each dimer
 [corresponding to ${\bf r}$ or $\mu$ in Eq.~(\ref{expansion})], and 
(ii) {\sl excited states in the transverse direction} 
for the relative motion of two dimers [corresponding to ${\bf R}$
or $\nu$ in Eq.~(\ref{expansion})].
Neglecting both types of closed-channel excitations, Eq.~(\ref{basiceq})
can be solved numerically for arbitrary $a_\perp/a$
as in Ref.~\cite{mora}.  
The result is shown in the inset of Fig.~\ref{f1}.
In addition, this approximation allows to extract $a_{dd}$ in both
limits analytically: in the dimer limit,
we find $a_{dd}= - \kappa_0 a_\perp^2/(2a) + 2\kappa_1 a,$
where $\kappa_0=1/4$ and $\kappa_1\simeq 0.319$,
while in the BCS limit, $a_{dd} = \eta_0 a_\perp^2/|a|$ 
with $\eta_0\simeq 0.402$.  The exact (numerical) result
for arbitrary $a_\perp/a$ agrees to within $\pm 0.05$ in $a_{dd}/a_\perp$ 
with a simple interpolation formula obtained by simply adding
these two limiting results.  For practical purposes, the interpolation
is therefore virtually exact.
Let us then turn to the effects of closed-channel excitations. 
In the BCS limit, excitations of type (ii) 
are irrelevant \cite{mora}, but type-(i) excitations 
are important.  Their inclusion results in the exact value $\eta_0=1/4$,
see Eq.~(\ref{tfres}), which also follows from the
solution of Eq.~(\ref{basiceq}) including type-(i) excitations \cite{mora3}. 
In the dimer limit, inclusion of the closed channels leads to the 
correct value $\kappa_0\approx 0.83$, see Eq.~(\ref{dimerlimit}). 
Incidentally, the two excitation types can be 
disentangled \cite{mora3}, and we find $a_{dd}^{3D}\approx 0.66a$ by 
just neglecting type-(i) excitations, which is 
already close to the exact value
$a_{dd}^{3D}\approx 0.6a$ \cite{petrov04}.
Type-(ii) excitations are obviously important in the dimer
limit, which may be valuable input for diagrammatics 
\cite{tokatly,strinati2}.  The exact limiting results for $a_{dd}$ 
are shown in the main part of Fig.~\ref{f1} 
as dashed curves.  For the full crossover, the additive interpolation formula
is again expected to be highly accurate. Notably, this
predicts $a_{dd}=0$ for $\Omega_B\approx 0.3$. At this point,
a {\sl CIR for dimer-dimer scattering} occurs, see Eq.~(\ref{1dd}),
where the interaction strength $g_{dd}$ diverges and changes sign.
Interestingly, the dimer-dimer CIR takes place at a different value for 
$\Omega_B$ (and hence $a_\perp/a$) 
than the atom-atom CIR.

\begin{figure}[t!]
\scalebox{0.3}{
\includegraphics{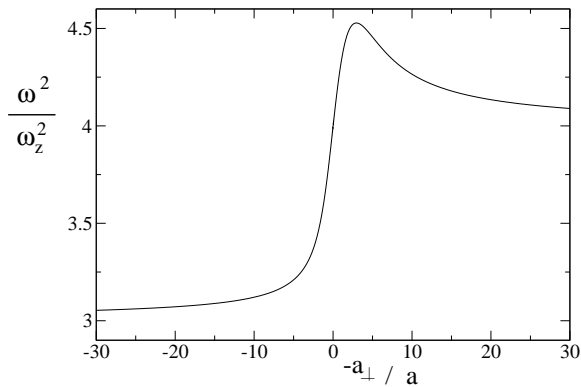} }
\caption{\label{f2}
Squared ratio of breathing and dipole mode frequency as a 
function of $-a_\perp/a$. Here we have chosen 
$N \omega_z / \omega_\perp = 1/3$.
}
\end{figure}

In experiments, quasi-$1$D regimes can be obtained
in arrays of very elongated traps with a shallow confinement in the 
longitudinal direction. Typical trap frequencies
are $\omega_\perp /2 \pi \approx 70$~kHz and $\omega_z /2 \pi \approx 250$~Hz,
with  $N\approx 100$ atoms per tube to ensure
the $1$D condition $N<\omega_\perp/\omega_z$ \cite{esslinger}. 
The BCS-BEC crossover can be investigated using a Feshbach resonance,
which leads to changes in the density profile \cite{tokatly},
excitation gaps \cite{fuchs} and ground state energy that can be probed via
release energy \cite{bourdel} and rf spectroscopy 
measurements \cite{chin,esslinger}.
A probably more precise approach is to measure collective axial modes.
The dipole mode frequency is always $\omega_z$, irrespective of interactions.
Using a sum rule approach \cite{astra}, we calculated the frequency of the
lowest compressional (breathing) mode from the mean-square size of the cloud
$\omega^2 = - 2 \left(  d \ln \langle z^2 \rangle/ d\omega_z^2\right)^{-1}$, 
see Fig.~\ref{f2}, by solving Eqs.~(\ref{lieb}) using our results for
 $a_{dd}$.  Limiting values are $\omega = \sqrt{3} \omega_z$ in the 
dimer limit,
and $\omega = 2 \omega_z$ both in the BCS limit and close to $a_{dd}=0$.
We hope that this prediction will soon be tested. - 
This work was supported by the DFG-SFB TR12.

\end{document}